\documentclass{article}

    \usepackage[preprint]{neurips_2025}

\usepackage[utf8]{inputenc} 
\usepackage[T1]{fontenc}    
\usepackage{hyperref}       
\usepackage{url}            
\usepackage{booktabs}       
\usepackage{amsfonts}       
\usepackage{nicefrac}       
\usepackage{microtype}      
\usepackage{xcolor}         
\usepackage{graphicx}
\usepackage{amsmath}
\usepackage{mdframed}
\usepackage{lipsum}
\usepackage{threeparttable}
\usepackage[english]{babel}
\usepackage{amsthm}
\usepackage{amssymb}
\newmdtheoremenv{theo}{Theorem}

\title{Train to Defend: First Defense Against Cryptanalytic Neural Network Parameter Extraction Attacks}
\author{%
  Ashley Kurian and Aydin Aysu \\
  Department of Electrical and Computer Engineering\\
  North Carolina State University\\
  \texttt{akurian@ncsu.edu, aaysu@ncsu.edu} \\
}

\begin{document}

\maketitle

\vspace{-1.5em}
\begin{abstract}
\vspace{-0.5em}
Neural networks are valuable intellectual property due to the significant computational cost, expert labor, and proprietary data involved in their development. Consequently, protecting their parameters is critical not only for maintaining a competitive advantage but also for enhancing the model's security and privacy. Prior works have demonstrated the growing capability of cryptanalytic attacks to scale to deeper models. In this paper, we present the first defense mechanism against cryptanalytic parameter extraction attacks. Our key insight is to eliminate the neuron uniqueness necessary for these attacks to succeed. We achieve this by a novel, extraction-aware training method. Specifically, we augment the standard loss function with an additional regularization term that minimizes the distance between neuron weights within a layer. Therefore, the proposed defense has zero area-delay overhead during inference. We evaluate the effectiveness of our approach in mitigating extraction attacks while analyzing the model accuracy across different architectures and datasets. When re-trained with the same model architecture, the results show that our defense incurs a marginal accuracy change of less than 1\% with the modified loss function. Moreover, we present a theoretical framework to quantify the success probability of the attack. When tested comprehensively with prior attack settings, our defense demonstrated empirical success for sustained periods of extraction, whereas unprotected networks are extracted between 14 minutes to 4 hours.
\vspace{-0.5em}
\end{abstract}
\vspace{-0.5em}

\section{Introduction}
\vspace{-0.75em}
\label{intro}
Neural networks hold significant value as their training process is computationally intensive, requires skilled labor, involves proprietary intellectual property, and demands extensive effort in dataset collection. Today, neural networks are ubiquitous, powering a wide range of applications due to their superior performance in various machine-learning tasks. The rise of machine learning as a service (MLaaS) has made it common for users to query models by providing input and receiving output, while model providers seek to obfuscate model parameters to protect their intellectual property. However, the increasing adoption of neural networks has also led to a rise in model-stealing attacks~\citep{tramer2016stealing,papernot2017practical,orekondy2019knockoff}, which aim to create \emph{high-fidelity}~\citep{jagielski2020high} replicas of the original models. These attacks target various components of the neural network, including hyperparameters~\cite{batina2018csi,deepem,kurian2025tpuxtract} (such as layer types and configurations) and learned parameters~\cite{carlini2020cryptanalytic,weightextract_sidechannel}(trained weights and biases). The significant cost and effort savings compared to training a model from scratch make model theft an appealing strategy for adversaries. Moreover, model extraction eases other attacks, such as membership inference~\cite{7958568,DBLP:journals/tches/ShuklaABMM23} and input poisoning~\cite{chen2017targeted}, further exacerbating security concerns.

Various attacks have demonstrated the feasibility of extracting neural network parameters through cryptanalytic techniques~\cite{carlini2020cryptanalytic, canales2024polynomial, shamir2023polynomial, foerster2024beyond}. These attacks treat parameter recovery as a structured mathematical problem, enabling the extraction of weights using a polynomial number of queries and within polynomial time with respect to model size on deep models~\cite{foerster2024beyond}. Despite their effectiveness, cryptanalytic extraction attacks remain a major threat that has not been addressed to date, with their capabilities growing rapidly.

In this paper, we conduct the first in-depth analysis of the cryptanalytic extraction attack and identify key parameter conditions required for its success. Specifically, we observe that extracting weight magnitudes depends on the uniqueness of neuron weights within a layer---the less similar the weights, the more effective the attack. To mitigate this, we propose a training strategy that enforces weight similarity among intra-layer neurons, inherently enhancing the model's resistance to such attacks. This is achieved by incorporating a weight similarity constraint into the training loss function alongside the standard loss, which minimizes prediction error. However, enforcing similarity among intra-layer neuron weights can lead to a change in model accuracy. To address this, we explore various optimization strategies to ensure that the accuracy change remains minimal while maintaining a strong defense against parameter extraction. Furthermore, we evaluate our countermeasure across different neural network configurations used in prior end-to-end attack~\cite{foerster2024beyond}, including various datasets, architectures, training, and extraction seeds. 

The main contributions of this paper include the following:
\begin{itemize}

\item We conduct a comprehensive analysis of cryptanalytic attacks, examining their underlying mechanisms and identifying the key parameter configurations that enhance their efficiency and likelihood of success. Understanding these crucial dependencies provides valuable insights into the fundamental weaknesses that adversaries exploit, laying the foundation for developing effective countermeasures against these attacks. 

\item In order to counteract cryptanalytic attacks, we propose an \emph{extraction-aware training} strategy. This approach involves introducing a regularization term into the loss function that intentionally adjusts the model's parameters during training. The goal is to encourage parameter configurations that minimize the effectiveness of cryptanalytic attacks, making it significantly difficult for adversaries to extract sensitive model information successfully. 

\item{A key advantage of our proposed defense is being zero-overhead at inference. The process is built into the training, and we show that re-training the same model hyperparameters with our technique can be sufficient to mitigate the attacks and has no run-time overheads.}

\item We analyze and implement various optimizations within our defense strategy to minimize the accuracy change caused by the inclusion of the defense mechanism while quantifying that the model remains robust and secure against cryptanalytic attacks. Our goal is to strike a balance between maintaining high model performance and enhancing resilience, thereby achieving an effective defense without compromising the model's accuracy.

\item We evaluate our defense strategy across various neural network configurations that were attacked in prior works, including different datasets, model depths, numbers of neurons per layer, and training seeds. This evaluation allows us to assess the robustness and effectiveness of our defense under diverse conditions and configurations, ensuring its generalizability and reliability across a range of practical scenarios demonstrated in prior works.

\item A major issue with prior attack evaluations was relying empirically on a set threshold (36 hours of attack runtime) for determining the attack success. To address this limitation, we propose a theoretical framework to assess attack success probability as a function of intra-layer neuron parameter similarity.
\end{itemize}

Our research demonstrates that extraction-aware training can effectively thwart cryptanalytic model parameter recovery attacks while preserving model accuracy with only a marginal change. By employing this lightweight defense strategy, we show that neural networks can be robustly protected from model stealing threats, offering a practical solution to safeguard sensitive model parameters without compromises in performance, memory footprint, or latency.


\vspace{-0.75em}
\section{Background}
\vspace{-0.75em}
\label{background}
\subsection{Related work and our solution}
\vspace{-0.5em}
Parameter extraction attacks on deep neural networks have been extensively studied over the past decades~\cite{fefferman1994reconstructing,Lowd2005AdversarialL,tramer2016stealing,milli2019model,rolnick2020reverse,martinelli2024expandandclusterparameterrecoveryneural}. These attacks seek to recover the weights and biases with sufficient numerical precision to achieve functional equivalence~\cite{jagielski2020high}, meaning the extracted model should produce identical predictions to the original model for any input. Cryptanalytical parameter extraction attack exploits the fact that ReLU neural networks are piecewise linear functions, and thus queries at the points where the input to the activation is zero reveal information about the model parameters. The seminal cryptanalytic extraction attack shows the recovery of normalized values of weights and biases (\emph{neuron signature}) with a polynomial number of queries with respect to model size~\cite{carlini2020cryptanalytic}. However, their method required an additional exhaustive search to recover the signs of the weights, thereby introducing exponential time complexity to the attack. As a result, their evaluation was limited to shallow neural networks with a maximum of three layers.
Martinez et al.~\cite{canales2024polynomial} addressed the sign extraction limitation by introducing techniques to recover neuron signs, enabling a fully polynomial-time attack. They demonstrated sign extraction on deeper networks with up to eight layers. Foerster et al.~\cite{foerster2024beyond} combined signature extraction and sign recovery to perform the first end-to-end parameter extraction attack. They introduced further optimizations to the sign extraction step, improving efficiency by 14.8×. These attacks assume access to confidence scores rather than just class labels for parameter extraction. More recently, cryptanalytic approaches have emerged that bypass this requirement and perform extraction using only the predicted labels~\cite{carlini2024polynomialtimecryptanalyticextraction,cryptoeprint:2024/1403}. 

The advancements in neural network parameter extraction have raised significant concerns about protecting proprietary model values. However, no countermeasures have been developed to defend against these attacks. We examine the key assumptions and parameter configurations that contribute to the success of these attacks. Through this analysis, we observe that a fundamental assumption in signature extraction is that the intra-layer neuron parameters must be unique~\cite{carlini2020cryptanalytic}. Moreover, certain neurons are classified as \emph{hard neurons}, i.e., neurons whose signs are recovered with low confidence~\cite{foerster2024beyond}. This occurs when it is impossible to craft an input perturbation that isolates the activity of the target neuron due to its similarity with other neurons. For such hard neurons in the previous layer, prior work propose recovering the next layer’s signature by brute-forcing their signs. Their method performs parallel signature recovery across sign combinations, selecting the one that yields correct signatures for the next layer. The approach remains feasible for up to ten low-confidence neurons, beyond which it becomes computationally prohibitive~\cite{foerster2024beyond}.

Our goal in this paper is to leverage these insights and incorporate conditions into the training phase to make neurons \emph{harder}. This is achieved by increasing intra-layer neuron parameter similarity, thereby making the model inherently resilient to cryptanalytic attacks. Specifically, we introduce a regularization term in the loss function that minimizes the distance between neuron parameters, while simultaneously optimizing for the primary learning task. While this approach strengthens model security, it impacts accuracy. To mitigate this trade-off, we explore several optimizations, including applying the defense only to the first layer, since protecting it can help safeguard the rest of the model in layer-wise attacks. We also evaluate strategies such as adjusting the strength of the regularization term to balance model accuracy and security. Finally, we assess the effectiveness and generalizability of our defense across different model architectures, datasets, training, and extraction seeds.
\vspace{-1.14em}
\subsection{Threat model}
\vspace{-0.5em}
\label{threat_model}
We adopt the standard assumptions commonly used in neural network cryptanalytic parameter extraction attacks~\cite{carlini2020cryptanalytic, canales2024polynomial,shamir2023polynomial,foerster2024beyond,carlini2024polynomialtimecryptanalyticextraction,cryptoeprint:2024/1403}. The target model is a fully connected neural network that uses ReLU activations in the hidden layers and a linear activation in the output layer. The attacker has full knowledge of the target model's hyperparameters. The attack operates with high precision, capable of recovering parameters up to 32-bit floating point precision. The attacker is also granted the ability to input arbitrary data and observe the corresponding class labels or logits scores, effectively treating the model as a black-box system with the objective of reconstructing a functionally equivalent replica of the original network. The model can support any number of output classes. Side-channel attacks~\cite{horvath2023barracuda,weightextract_sidechannel} are considered out of scope, as they rely on auxiliary information such as power consumption or electromagnetic emissions, which may not be accessible in all application scenarios.

\begin{figure}[t]
\centering
\vspace{-1.5em}
\includegraphics[width=1\textwidth]{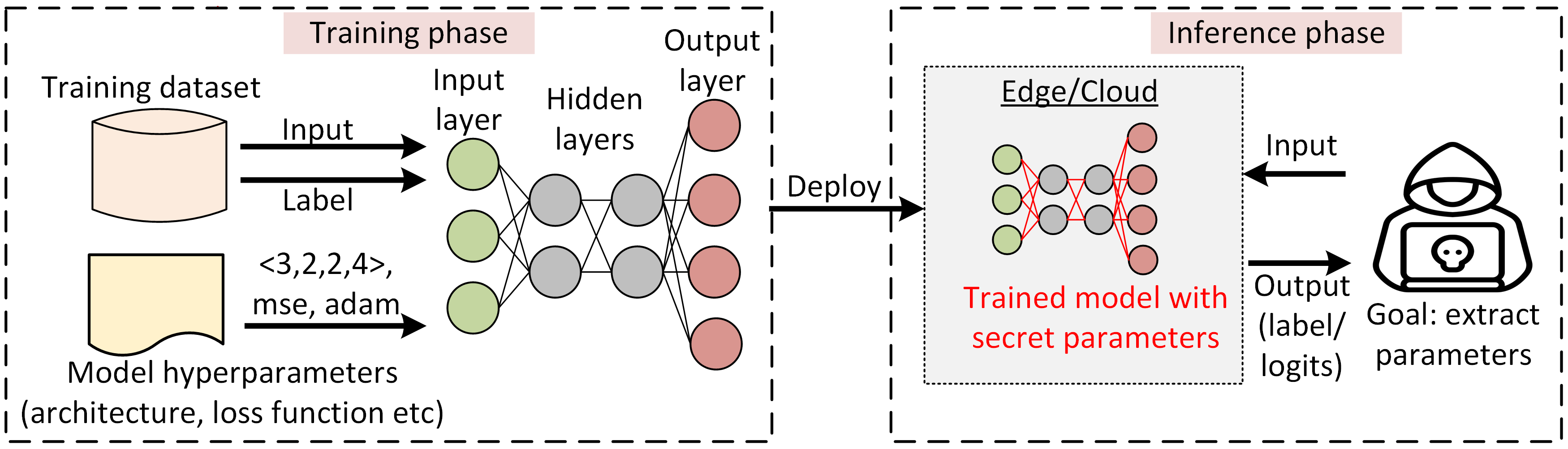}
\vspace{-1em}
\caption{Schematic depicting how a deployed neural network becomes vulnerable to attack. A neural network is trained on a proprietary dataset and later deployed for inference. During the inference phase, an attacker queries the model to extract labels or logit scores for a cryptanalytic attack, while the defender aims to protect the model's secret parameters.}
\label{Threat model}
\end{figure}

The overall attack setup is illustrated in Figure~\ref{Threat model}, which depicts the training and deployment phases of the model. As shown, the neural network is trained using a proprietary dataset and specific hyperparameters, including the model architecture---for example, three input nodes, two hidden layers with two nodes each, and four output classes <3,2,2,4>, with the mean squared error (MSE) loss function and the Adam optimizer. We follow the assumption that training is trusted. Once trained, the model, with its secret parameters, is deployed for inference. An attacker can then query the model to obtain class labels or logit scores to perform a cryptanalytic attack. The defender's objective is to secure the trained model parameters against such attacks.

The goal of the model provider is to build a robust neural network that remains resilient against cryptanalytic parameter extraction attacks at inference.
From the defender's perspective, we assume full control over the training process, including the ability to modify the loss function. 
We assume that training is trusted and not subject to model extraction attacks.  
Once deployed, the adversary can query the model and receive responses to execute cryptanalytic attacks during inference. 

\vspace{-0.5em}
\subsection{Cryptanalytic parameter extraction---A walkthrough}
\vspace{-0.5em}
\label{walkThrough}
This section walks through the signature extraction technique~\cite{carlini2020cryptanalytic} using a simple neural network for illustration. The attack recovers the weights of individual neurons by carefully probing how the output of the network changes in response to small input perturbations. The process begins by identifying critical points—--input values where a neuron's ReLU output is zero following the terminology used in prior works~\cite{carlini2020cryptanalytic,canales2024polynomial,foerster2024beyond}. Gradients are then computed around these points to capture directional information about the weights. By comparing the network's behavior when a neuron is active versus inactive, the attack isolates the influence of a single neuron and recovers its normalized parameters.

\begin{figure}[t]
\centering
\vspace{-1.8em}
\includegraphics[width=1\textwidth]{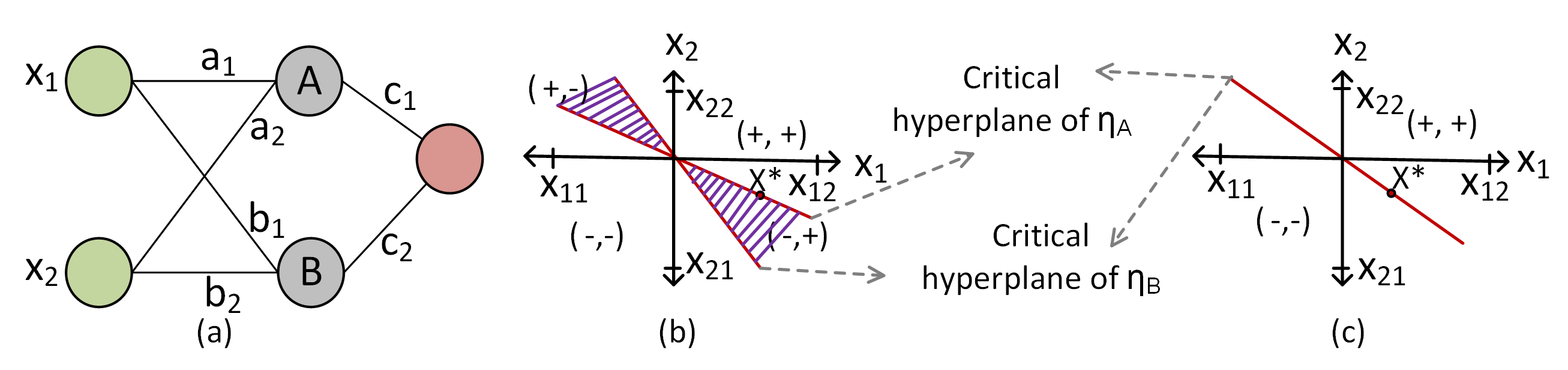}
\vspace{-2em}
\caption{(a) neural network schematic with <2,2,1> configuration (b) input space showing the critical hyperplanes of \(\eta_A\) and \(\eta_B\) with distinct parameters (c) overlapping critical hyperplanes of \(\eta_A\) and \(\eta_B\) when their parameters are identical.}
 \vspace{-1.1em}
\label{NN}
\end{figure}

 Consider a feedforward neural network with architecture ⟨2, 2, 1⟩, as shown in Figure \ref{NN}(a), where neuron \emph{A} (\(\eta_A\)) has weights \((a_1, a_2)\), neuron \emph{B} (\(\eta_B\)) has weights \((b_1, b_2)\) and the input vector \emph{X}= \((x_1, x_2)\). A \emph{k}-deep neural network \(f_\theta(x)\), parameterized by \(\theta\) maps input space \emph{X} to output space \emph{Y}. With ReLU activation \(\sigma\), the output is:
 \begin{equation}
 f_\theta(x) = c_1\sigma(a_1 x_1 + a_2 x_2) + c_2\sigma(b_1 x_1 + b_2 x_2)
 \end{equation}

Figure \ref{NN}(b) shows the input space and corresponding critical hyperplanes of \(\eta_A\) and \(\eta_B\), which represent the plane containing points where at least one of the ReLU inputs becomes zero.
The critical points lie on the critical hyperplane. To recover the first-layer parameters, the attack traces the gradient flow from input coordinate \emph{i} through the target neuron to the output. At each critical point \(x^*\), the attack computes the directional derivatives of\(f_\theta(x)\) by applying small positive and negative perturbations \(\epsilon\) along the input axis \(e_i\).
\begin{equation}
   \alpha_i^{+}=\frac{\partial f(x^* + \varepsilon e_i)}{\partial e_i}, \alpha_i^{-} = \frac{\partial f(x^* - \varepsilon e_i)}{\partial e_i}
\end{equation}
where \(e_i \in \mathbb{R}^{d_0 = 2}\), \(d_0\) is the number of input neurons. The signature can then be recovered by computing (\(\alpha_i^{+}+\alpha_i^{-}\))~\cite{carlini2020cryptanalytic}. Assume we want to recover parameters of \(\eta_A\) using a critical point \(x^*\). Suppose \(x^* + \varepsilon e_i\) lies in a region where both \(\eta_A\) and \(\eta_B\) are active (represented using (+,+)), and \(x^* - \varepsilon e_i\) lies in a region where \(\eta_A\) is inactive and \(\eta_B\) is active (represented using (-,+)). In this case:
\begin{equation}
    f_\theta(x^* + \varepsilon e_1) =c_1(a_1 (x_1+\varepsilon e_1) + a_2 x_2) + c_2(b_1 (x_1+\varepsilon e_1) + b_2 x_2)
\end{equation}
 \begin{equation}
    f_\theta(x^* - \varepsilon e_1) = c_1\cdot0 + c_2(b_1 (x_1-\varepsilon e_1) + b_2 x_2)
 \end{equation}
The gradients are:
\(\alpha_1^{+} = c_1a_1\varepsilon + c_2b_1\varepsilon\) and \(\alpha_1^{-}= -c_2b_1\varepsilon\). Their sum $\alpha_1^{+}+\alpha_1^{-} = c_1a_1\varepsilon$, isolates the first weight coordinate \(a_1\) of \(\eta_A\). Similarly, for direction 2, the sum \(\alpha_2^{+}+\alpha_2^{-} = c_1a_2\varepsilon\) reveals the second weight coordinate \(a_2\). The normalized signature becomes:
\begin{equation}
  (\frac{\alpha_1^{+} + \alpha_1^{-}}{\alpha_1^{+} + \alpha_1^{-}},\ 
\frac{\alpha_2^{+} + \alpha_2^{-}}{\alpha_1^{+} + \alpha_1^{-}}) = (\frac{a_1}{a_1},\ \frac{a_2}{a_1})  
\end{equation}
This recovers the normalized weights of $\eta_A$. Notably, terms from other neurons cancel out (in this case, that of $\eta_B$), enabling isolated recovery of the target neuron's parameters. Repeating the process with a critical point for $\eta_B$ allows recovery of its signature as well. See~\cite{carlini2020cryptanalytic,canales2024polynomial,shamir2023polynomial} for further details on signature and sign recovery. Attacks that rely on predicted labels adopt a similar strategy but begin by identifying critical hyperplanes through the detection of transition points---input values where the model’s predicted class changes, thereby revealing the boundaries between decision regions~\cite{carlini2024polynomialtimecryptanalyticextraction}.

\vspace{-0.9em}
\section{Proposed defense against cryptanalytic parameter extraction}
\vspace{-0.5em}
\label{proposedWork}
We propose the first countermeasure against cryptanalytic parameter extraction attacks. These attacks can target deep models trained on real-world datasets, requiring only black-box access in polynomial time and with a polynomial number of queries. Our defense addresses the main assumption underlying the success of these cryptanalytic attacks. The signature and sign extraction process is based on the distinctness of the neuron parameters, with the success rate of the attack increasing as the uniqueness between the neuron parameters increases. Our defense challenges this assumption by promoting similarity among neuron parameters during the training phase. This is achieved by adding a regularization term to the loss function that reduces the distance between neurons while still optimizing for the original loss function. 


\vspace{-0.75em}
\subsection{Why cryptanalytic attack fails by making neurons similar?}
\vspace{-0.5em}
\label{defense_walkthrough}
In this section, we provide a mathematical explanation of how making neuron parameters similar disrupts the cryptanalytic attack procedure, using the same example from Section \ref{walkThrough}. Consider the neural network shown in Figure \ref{NN}(a). If the parameters of neurons $\eta_A$ and $\eta_B$ are the same—that is,
\((a_1, a_2)=(b_1, b_2)\)—their critical hyperplanes overlap. As a result, both neurons will be in the same state (either active or inactive) for any given input, as shown in Figure \ref{NN}(c). Suppose we are trying to extract the signature of $\eta_A$ and let $x^*$ be a critical point. Assume that the input $x^* + \varepsilon e_i$ activates both neurons and  the input $x^* - \varepsilon e_i$ deactivates both. Then, for direction 1: 
\begin{equation}
    f_\theta(x^* + \varepsilon e_1) =c_1(a_1 (x_1+\varepsilon e_1) + a_2 x_2) + c_2(b_1 (x_1+\varepsilon e_1) + b_2 x_2)
\end{equation}
\begin{equation}
    f_\theta(x^* - \varepsilon e_1) = c_1\cdot0 + c_2\cdot0
\end{equation}
The resulting gradients are: $\alpha_1^{+} = c_1a_1\varepsilon + c_2b_1\varepsilon$ and $\alpha_1^{-}= 0$. The sum becomes: $\alpha_1^{+}+\alpha_1^{-} = c_1a_1\varepsilon + c_2b_1 \varepsilon$. This result includes the influence of both $\eta_A$ and $\eta_B$, making it impossible to isolate the contribution of $\eta_A$ for any given input perturbation. Therefore, the attack fails to recover the correct signature. Similarly, for direction 2: $\alpha_2^{+}+\alpha_2^{-} = c_1a_2\varepsilon + c_2b_2\varepsilon$. The normalized signature is:
\begin{equation}
  (\frac{\alpha_1^{+} + \alpha_1^{-}}{\alpha_1^{+} + \alpha_1^{-}},\ 
\frac{\alpha_2^{+} + \alpha_2^{-}}{\alpha_1^{+} + \alpha_1^{-}}) = (1,\ \frac{ c_1a_2\varepsilon + c_2b_2\varepsilon}{c_1a_1\varepsilon + c_2b_1 \varepsilon})
\end{equation}
This signature is incorrect because it mixes contributions from both neurons. The correct signature should be $(1,a_2/a_1)$, which reflects the behavior of only the target neuron (as derived in Section~\ref{walkThrough}). Making neuron parameters similar causes them to activate or deactivate together for any input, preventing the attack from isolating a specific neuron and leading to its extraction.
\vspace{-0.5em}
\subsection{Probability that a random input fails defense}
\vspace{-0.5em}
\label{theory_probability}
A key limitation of prior attack evaluations is their reliance on an arbitrary empirical threshold (e.g., 36 hours of extraction) to define success. To address this, we introduce a theoretical framework that calculates the attack success probability as a function of the intra-layer neuron weight similarity. 
The key insight is that an attack fails when neurons are in the same state (either active or inactive) for any input, as it is not possible to isolate target neuron activity. However, if the parameters of the neurons differ, their critical hyperplanes diverge, and the attack succeeds. We quantify the probability of this success by calculating the area between the hyperplanes, where the neurons' states differ. This gives the probability of successfully extracting a single neuron pair. The total attack success probability is then obtained by identifying the pair of neurons with the greatest dissimilarity, representing the worst-case scenario. 
\begin{theo}
The probability of extracting parameters of a neuron pair is the ratio between the measure of the input region lying between their corresponding hyperplanes and the total measure of the entire input space. This probability is is upper bounded by \(\frac{|\!K_1| + |\!K_2| + \dots + |\!K_{N-1}|}{2} \cdot 
\frac{x_{12}^2 + x_{11}^2}{(x_{12} - x_{11})^2}\) where\(\quad K_i = \frac{-a_i}{a_N} - \frac{-(a_i + \delta_i)}{a_N + \delta_N}, \quad \text{for } i = 1, 2, \dots, N-1\);\ \(N\) denotes the input dimension, \(a_i\) refers to the \(i\)-th parameter of the neuron, \(\delta_i\) is the \(i\)-th parameter difference between the neuron pair, and  \  \(x_{11}\) and \(x_{12}\) are respectively the lower and upper limits of an input dimension.


\end{theo}
\renewcommand\qedsymbol{$\blacksquare$}
\begin{proof}
Consider the neural network in Figure \ref{NN}(a). If the parameters of neurons \(\eta_A\) and \(\eta_B\) are identical, their critical hyperplanes overlap, as shown in Figure \ref{NN}(c). In that case, both neurons always behave the same, and the defense works perfectly. However, when their parameters differ by \((\delta_1, \delta_2)\)—meaning \(b_1=a_1 + \delta_1\) and \(b_2=a_2 + \delta_2\)—their critical hyperplanes do not align. The equations for their critical hyperplanes are:
\begin{equation}
    \text{For}\ \eta_A: a_1x_1 + a_2x_2=0
\end{equation}
\begin{equation}
  \text{For}\ \eta_B: b_1x_1 + b_2x_2=0 \implies (a_1 + \delta_1)x_1 + (a_2 + \delta_2)x_2=0 
\end{equation}
The shaded region in Figure \ref{NN}(b) shows the area between these two hyperplanes, where the neurons behave differently. Inputs that fall in this area cause the attack to succeed. Let \(x_{11}\) and \(x_{12}\) be the lower and upper limits for the input \(x_1\) and \(x_{21}\) and \(x_{22}\) for \(x_2\). For INT8 inputs, these range from -128 to +127.
For simplicity, we use \(x_{11}\) and \(x_{12}\) as the common lower and upper bounds for all input dimensions in the derivation, as each dimension shares the same input range.
\begin{equation}
    \text{Area between hyperplanes of} \ \eta_A \ \text{and} \ \eta_B= \int_{x_{11}}^{x_{12}} \left| \left( \frac{-a_1}{a_2} \cdot x_1 \right) - \left( \frac{-(a_1 + \delta_1)}{(a_2 + \delta_2)} \cdot x_1 \right) \right| \, dx_1
\end{equation}

\begin{equation}
    = \int_{x_{11}}^{x_{12}} |K \cdot x_1| \, dx_1 
= \frac{|K|}{2} \left(x_{12}^2 + x_{11}^2\right)
\ \text{where}, K = -\frac{a_1}{a_2} + \frac{a_1 + \delta_1}{a_2 + \delta_2}
\end{equation}
\begin{equation}
   \text{Total area of the input space} = \int_{x_{11}}^{x_{12}} \int_{x_{11}}^{x_{12}} dx_2\, dx_1 = (x_{12} - x_{11})(x_{12} - x_{11})
\end{equation}
\begin{equation}
   \text{Attack success probability} = \frac{\text{Area between hyperplanes}}{\text{Total area}}=\frac{|K|}{2} \cdot \frac{x_{12}^2 + x_{11}^2}{(x_{12} - x_{11})^2}
\end{equation}
This is the probability that a randomly chosen input will fall between the hyperplanes, meaning the neurons respond differently, and the attack succeeds. As the differences \(\delta_1\) and \(\delta_2\) increase, the two hyperplanes move further apart, increasing this probability. For an N-dimensional input space, due to the triangle inequality, the probability is upper bounded and can be computed as:
\begin{equation}
    \text{Probability of attack success, \emph{P}} \leq 
\frac{|\!K_1| + |\!K_2| + \dots + |\!K_{N-1}|}{2} \cdot 
\frac{x_{12}^2 + x_{11}^2}{(x_{12} - x_{11})^2}
\end{equation}
   \[\text{where} \quad K_i = \left( \frac{-a_i}{a_N} \right) - \left( \frac{-(a_i + \delta_i)}{a_N + \delta_N} \right), \quad \text{for } i = 1, 2, \dots, N-1 \qedhere \]
\end{proof}
\renewcommand\qedsymbol{QED}
This derivation assumes that the normal vectors of the two neurons’ hyperplanes point in the same direction. In this case, the area between the hyperplanes (the shaded region in Figure \ref{NN}(b)) represents where the neuron states differ, and thus where the attack succeeds. However, if the normal vectors point in opposite directions, then the region where the neurons differ in state flips---it becomes the region where they previously had the same state. In other words, the failure region becomes the unshaded area in Figure \ref{NN}(b). So, in this case, the probability of attack success becomes (\(1-P\)). The direction of the normal vectors is captured by the angle \(\theta\) between them:
\begin{equation}
\theta = \cos^{-1}\left( \frac{\vec{a} \cdot \vec{b}}{\|\vec{a}\| \|\vec{b}\|} \right) 
\quad \text{where } \vec{a} = (a_1, a_2, \dots, a_N),\ \vec{b} = (b_1, b_2, \dots, b_N)
\end{equation}
\begin{equation}
\text{Probability of attack success} =
\begin{cases}
P & \text{if } 0 \leq \theta \leq \frac{\pi}{2} \quad (\text{same direction}) \\
1 - P & \text{if } \frac{\pi}{2} < \theta \leq \pi \quad (\text{opposite direction})
\end{cases}
\end{equation}
\vspace{-0.9em}
\subsection{Extraction-aware neural network training}
\vspace{-0.5em}
\label{extraction_aware_training}
Adding a term to the loss function to increase similarity between neuron parameters can help defend against cryptanalytic attacks. We introduce a modified loss function:
\begin{equation}
    \text{Loss function} = original\_loss + \lambda_{similarity} * total\_similarity\_loss
\end{equation}
Here, $original\_loss$ refers to the standard function (such as MSE) that reduces the error between predicted and true labels. We introduce the additional term ($\lambda_{similarity} * total\_similarity\_loss$) to improve the security of the model. The factor $\lambda_{similarity}$ controls the impact of the similarity term, allowing a balance between accuracy and security—--higher values increase security but may reduce accuracy. The $total\_similarity\_loss$ captures the overall similarity between neurons across all layers in the network: 
\begin{equation}
total\_similarity\_loss = \Sigma_{\text{layer}=1}^{\text{number of layers}} layer\_wise\_similarity\_loss
\end{equation}
\begin{equation}
    layer\_wise\_similarity\_loss = \Sigma_{i=1}^{(\text{number of neurons} - 1)} (p_i - p_{i+1})^2
\end{equation}
where \(p_i\) is the parameter of neuron \(i\). The $layer\_wise\_similarity\_loss$ 
computes the pairwise difference between parameters of neurons within the same layer. This term uses L2 regularization to penalize large differences, making neurons more similar.

\begin{figure}[bt]
\centering
\vspace{-1.2em}
\includegraphics[width=0.9\textwidth]{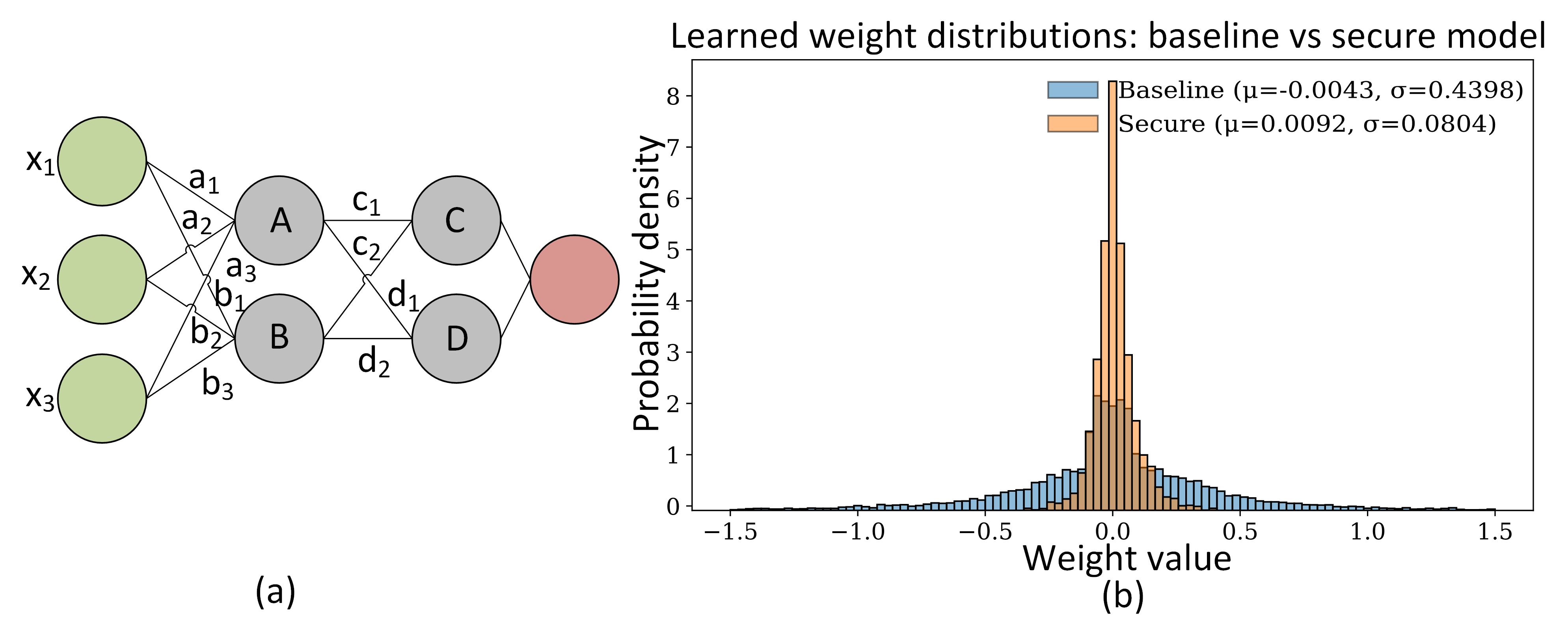}
\vspace{-1.5em}
\caption{(a) neural network schematic with <3,2,2,1> configuration (b) comparison of weight distributions in the first dense layer of the secure and baseline models. The secure model shows a more concentrated distribution centered around zero with lower variance (0.08), while the baseline model exhibits a wider spread (variance 0.43). Only the range [–1.5, +1.5] is shown, which contains approximately 90\% of the weights.}
 \vspace{-1.5em}
\label{NNa}
\end{figure}

To illustrate, consider a neural network with architecture <3, 2, 2, 1> as shown in Figure~\ref{NNa}(a). The $total\_similarity\_loss$ is the sum of the similarity losses for hidden layers 1 and 2:
\begin{equation*}
    total\_similarity\_loss=layer1\_loss + layer2\_loss
\end{equation*}
\begin{equation*}
 layer1\_loss= (a_1 - b_1)^{2} + (a_2 - b_2)^{2} + (a_3 - b_3)^{2}\; \text{and}\;    layer2\_loss = (c_1 - d_1)^{2} + (c_2 - d_2)^{2}  
\end{equation*}
In this example, the loss in each layer is computed based on the difference between the parameters of two neurons within that layer. For a layer with $n$ neurons, there will be $\frac{n(n-1)}{2}$ pairwise comparisons. The goal is to make neurons behave similarly so that their outputs are the same for any input. This similarity prevents the attacker from isolating and targeting a specific neuron, as multiple neurons will respond in the same way.
\vspace{-0.8em}
\subsection{Tuning our defense}
\vspace{-0.5em}
The additional term $\lambda_{similarity} * \ total\_similarity\_loss$ is introduced in the loss function to enhance model security. However, this regularization must be carefully controlled to avoid compromising model accuracy. To mitigate accuracy degradation, we propose the following tuning strategies:
\vspace{-0.5em}
\begin{itemize}
    \item Adjusting the regularization strength: By varying the impact factor $\lambda_{\text{similarity}}$, it is possible to control influence on accuracy while still making the model robust against attacks.
    \vspace{-0.25em}
    \item Restricting defense to the first layer: Since the attack proceeds layer by layer, securing only the first layer can be sufficient. If the parameters of the first layer are protected, an attack on the subsequent layers could be automatically prevented.
    \vspace{-0.25em}

    \item Defending a subset of neuron pairs: Instead of enforcing similarity across all neurons in the first layer, the defense can target only a percentage of the neuron pairs. For example, a layer with eight neurons results in 28 pairwise similarity terms (i.e., $\frac{8(8-1)}{2}$). Defending 50\% of the neurons involves enforcing similarity on 14 randomly selected pairs.
    \vspace{-0.5em}

\end{itemize}
These optimization strategies help balance security and accuracy, ensuring that the defense mechanism does not adversely impact the model’s performance.
\vspace{-1 em}
\vspace{-0.1em}
\section{Results}
\label{results}
\vspace{-1em}
We evaluate our defense by measuring the model accuracy and attack run-time for parameter extraction. The models are trained using the modified loss function discussed in Section~\ref{extraction_aware_training} to defend cryptanalytic parameter recovery attacks. To limit accuracy change, we apply optimizations such as restricting changes to the first layer parameters and adjusting $\lambda_{similarity}$. Table~\ref{table_results} shows results across different architectures, datasets, and training seeds. We follow the notation used in prior work~\cite{foerster2024beyond}. For example, MNIST784\_8x2\_1(s2) model refers to an MNIST-trained network with an input layer of size 784, two hidden layers, each of size 8, and an output layer of size 1. We compare secure models to their baseline by reporting accuracy, attack runtime, and $\lambda_{similarity}$ that makes them secure. All training and extraction were done on a high-performance cluster with Intel's Xeon processors comprising 400 compute nodes and over 14,000 cores. On average, each node is provisioned with 128 GB of memory. We take our models from the papers that demonstrate attacks~\cite{foerster2024beyond}, and we use exactly all the hyperparameters evaluated in those papers. Models are trained on random data and MNIST, using two different training seeds (s1 = 42, s2 = 10). The extraction times for signatures and signs, number of queries, are reported as mean and variance across four extraction seeds (0, 10, 20, and 30). We consider a model to be secure if the attack fails after extended periods of extraction\footnote{Prior work~\cite{foerster2024beyond} uses 36 hours, whereas we use 48 hours as the extraction time limit.}. For example, the model MNIST784\_16x8\_1(s2), which was previously compromised in 3.5 hours, becomes secure with a $\lambda_{similarity}$ of $10^{-9}$, while its accuracy drops by 0.92\%. In some cases, the secure model even performs better. For instance, the MNIST784\_16x3\_1(s1) model shows a 0.43\% increase in accuracy compared to its baseline version. The results show that for all the previously studied attack settings, our defense can secure the models through extraction-aware training with an accuracy change of less than 1\% with zero-overhead inference. 

Using the theoretical framework described in Section~\ref{theory_probability}, we first compute the probability of attack success for models protected using our defense strategy, and then we empirically estimate the corresponding extraction time based on this probability. For example, the attack success probability for the MNIST784\_16x8\_1(s2) model is calculated based on the neuron pair in the first layer with the greatest parameter dissimilarity. The attack success probability drops from 0.74 for the baseline to 0.0017 for the secure version, a $435 \times$ decrease. Figure~\ref{NNa}(b) shows the weight distributions for the secure and baseline models. The baseline model's weights are spread across a wider range of [–2.5, +2.5], while the secure model’s weights fall within [–0.2, +0.2] (Figure~\ref{NNa}(b) shows only a range of [–1.5, +1.5], where 90\% of the weights fall). To quantify this spread, the secure model has a lower weight variance of 0.08 compared to 0.43 in the baseline model, with weights more tightly clustered around zero. This indicates that the parameters in the secure model are more similar to each other, reducing the attack’s ability to isolate neurons and extract their parameters individually. 
\vspace{-0.8em}


\begin{table}
\vspace{-0.8em}
  \caption{Cryptanalytic model parameter extraction performance}
  \label{table_results}
  \centering
   \begin{threeparttable}
  \resizebox{\textwidth}{!}{%
  \begin{tabular}{lllcccccccc}
  
    \toprule
    \multicolumn{3}{c}{\textbf{Model Information}} & \multicolumn{2}{c}{\textbf{Signature [s]}} & \multicolumn{2}{c}{\textbf{Sign [s]}} & \multicolumn{2}{c}{\textbf{Queries}} &\multicolumn{2}{c}{\textbf{Secure Model Information}} \\
    \cmidrule(r){1-3} \cmidrule(r){4-5} \cmidrule(r){6-7} \cmidrule(r){8-9} \cmidrule(r){10-11}
    \textbf{Model (Training Seed)} &\textbf{ Accuracy} & \textbf{Params} & \textbf{Mean} & \textbf{Var} & \textbf{Mean} & \textbf{Var} & \textbf{Mean} & \textbf{Var} & \textbf{Accuracy change (\%)} & $\boldsymbol{\lambda_{similarity}}$  \\
    \midrule
    Random784-8x2-1 (s1) & NA\tnote{*}& 6280 & $9.19\cdot 10^2$& $9.6 \cdot 10^2$ &54.69  &36.6  &$6.55\cdot 10^5$ &$2.53 \cdot 10^9$ & NA & $10^{-5}$ \\
    Random784-16x2-1 (s1) & NA & 12560 & $1.37\cdot 10^3$ &$ 7.11 \cdot 10^2$&$86.67$  &  46.99& $1.31\cdot 10^6$& $2.46\cdot 10^7 $&  NA&  $10^{-4}$\\
    Random784-32x2-1 (s1) & NA & 25120 & $2.54\cdot 10^3 $&$4.29\cdot 10^3 $ & $1.64\cdot 10^2$ &$ 1.7\cdot 10^2 $&$2.62\cdot 10^6 $&$1.41\cdot 10^6 $& NA & $10^{-4}$ \\
    Random784-64x2-1 (s1) & NA & 50240 & $5.43\cdot 10^3$ &$ 2.81\cdot 10^5$ & $2.78\cdot 10^2$ & $1.06\cdot 10^3$ &$5.24\cdot 10^6$ &$1.66\cdot 10^{11}$ &  NA& $10^{-1}$ \\
    Random784-128x2-1 (s1) & NA & 100480 &$1.16\cdot 10^4$ & $7.61\cdot 10^5$ & $4.79\cdot 10^2$ & $1.07\cdot 10^3$ & $1.04\cdot 10^7$& $6.57\cdot 10^{11}$& NA & $10^{-1}$\\
    Random784-128x2-1 (s2) & NA& 100480 & $1.38\cdot 10^4$ & $9.88\cdot 10^4$ & $5.54\cdot 10^2$ &8.12  & $1.2\cdot 10^7$& $3.2\cdot 10^6$& NA & $10^{-1}$\\
     &  &  &  &  &  &  & & &  &  \\
     MNIST784-8x2-1 (s2)& 70.34 & 6280 & $7.87\cdot 10^2$ & $5.74\cdot 10^3$ & 48.54 &26.52  &$6.55\cdot 10^5$ &$7.74\cdot 10^3$ & $-0.32$ &$10^{-9}$  \\
      MNIST784-16x2-1 (s2)& 74.40 & 12560 & $1.3\cdot 10^3 $& $3.93\cdot 10^3$ &95.71  &0.29  &$1.16\cdot 10^6$ &$2.27\cdot 10^9 $& $+0.74$ & $10^{-9}$ \\
      MNIST784-32x2-1 (s2)& 86.37 & 25120 &  $2.68\cdot 10^3$& $5.48\cdot 10^4$ &$1.91\cdot 10^2$  & $2.81\cdot 10^2$  &$2.42\cdot 10^6$  &$2.8\cdot 10^7$ & $+0.83$ &  $10^{-7}$\\
      MNIST784-64x2-1 (s2)& 91.41 & 50240 & $6.19\cdot 10^{3}$ & $1.56\cdot 10^{5}$  &  $3.54\cdot 10^{2}$&$4.96\cdot 10^{3}$  &$5.45\cdot 10^{6}$ & $4.06\cdot 10^{10}$& $-0.37$ & $10^{-6}$ \\
      MNIST784-64x2-1 (s1)& 91.48 & 50240 & $5.41\cdot 10^3$  & $5.28\cdot 10^5$  & $3.97\cdot 10^2$  & $2.34\cdot 10^3$  & $4.44\cdot 10^6$ & $5.93\cdot 10^6$ & $-0.31$ &  $10^{-6}$\\
      &  &  &  &  &  &  & & &  &  \\
      MNIST784-16x8-1 (s2)& 88.67 & 12560 &$ 1.12\cdot 10^4$& $1.95\cdot 10^6$ & 80.37 &0.41  & $5.65\cdot 10^6$& $8.91\cdot 10^9$& $-0.92$ & $10^{-9}$ \\
      MNIST784-16x3-1 (s1)& 84.93 & 12560 & $2.65\cdot 10^3$  & $1.99\cdot 10^3$  & $1.04\cdot 10^2$  & 81.09 &$1.91\cdot 10^6$  & $1.15\cdot 10^5$ & $+0.43$ & $10^{-9}$ \\
      &  &  &  &  &  &  & & &  &  \\
    \bottomrule
    
  \end{tabular}
  }
  \begin{tablenotes}\scriptsize
\item[*]NA: Accuracy not reported for random models as parameters are trained on random data

  \vspace{-2.5em}
\end{tablenotes}
   \end{threeparttable}
\end{table}

\vspace{-0.5em}
\section{Discussion}
\vspace{-0.9em}
\label{discussion}
\subsection{Where does the attack struggle for secure models?}
\vspace{-0.5em}
The cryptanalytic parameter extraction on models secured by our defense is unsuccessful even after running the attack longer than the set threshold in prior papers~\cite{foerster2024beyond}. The attack works by gathering critical points for each neuron to compute its signature. It then applies graph clustering to group critical points based on signature similarity. For the attack to succeed, each neuron should have a unique signature\footnote{\emph{Zero weights} are excluded from this condition because neuron output remains constant regardless of the input.}; otherwise, it cannot specifically target individual neurons, as discussed in Section~\ref{defense_walkthrough}. The attack continues clustering until the number of clusters matches the number of neurons in the target layer. However, in secure models, neuron similarity leads to fewer clusters than expected, as some neurons share the same signature. The attack then collects more critical points by exploring multiple directions and repeating clustering. This process fails, which results in attack failure even after extended runtime. With our defense, the attack is still polynomial in time and number of queries. Our defense does not improve the theoretical security, but in practice, it increases the search complexity of the critical points per neuron.
\vspace{-0.8em}
\subsection{Model extraction with partial parameter recovery}
\vspace{-0.65em}
A question is whether an attacker can still recover the model by extracting only a few neurons in the first layer, or by randomly initializing the first layer and continuing the attack on subsequent layers. The answer is no. The attack on the following layers depends on having the correct output from the first layer, as it solves a system of equations based on those outputs. If the first layer is incorrect, the computed outputs are also wrong, leading to incorrect parameter recovery in the second layer. Worse, the errors accumulate with each deeper layer, causing the extraction to fail. Therefore, protecting only the first layer is enough to secure the entire model.
Moreover, our proposed defense is not fundamentally limited to the first layer. It may be extended to future layers for other possible attacks that are not seeking \emph{exact} high-fidelity capture with cryptanalysis.
\vspace{-0.8em}
\subsection{Limitations}
\vspace{-0.65em}
\label{limitations}
Our effort to secure models against cryptanalytic parameter extraction attacks comes at the expense of a marginal change (less than 1\%) in accuracy. Additionally, as cryptanalytic attacks have not yet been demonstrated on CNNs or LLMs, our analysis focuses on the MLP network architectures evaluated in prior works. At present, these attacks apply only to MLPs with piecewise linear activations. These are limitations of the current attack (which is rapidly evolving), not of the defense. The defense itself has no such limitations and can scale as attacks advance, since it disrupts the key assumption underlying the attack--—namely, neuron dissimilarity.
\vspace{-0.8em}
\subsection{Other possible defenses}
\vspace{-0.55em}
Other countermeasures such as parameter encapsulation, neural structure obfuscation, and injection of shortcuts or extra layers~\cite{zhou2023modelobfuscator} can be employed to protect models. However, these approaches introduce computational overhead and increase the size of the ML library. Additional defenses against query-based model extraction reduce the effectiveness of such attacks through various strategies, including adding deceptive neurons~\cite{szentannai2020preventing}, perturbing predictions by maximizing angular deviation between gradients to poison the attacker's training objective~\cite{orekondy2020predictionpoisoningdefensesdnn}, or reprogramming the adversary with arbitrary behavior~\cite{mazeika2022steeradversarytargetedefficient}, but incur run-time or memory overhead. By contrast, our defense provides protection with zero overhead during inference.

\vspace{-1 em}
\section{Conclusion}
\vspace{-0.65em}
\label{conclusion}
Neural networks are valuable intellectual property that are vulnerable to parameter extraction attacks capable of recovering model parameters using cryptanalytic techniques in polynomial time and with a polynomial number of queries. In this paper, we present the first defense against such attacks. Our approach introduces an extraction-aware training method that adds a regularization term to the standard loss function. This term increases similarity between neuron parameters, making it difficult for the attack to isolate and recover individual neurons. The defense adds zero overhead during inference. We evaluate it across different network architectures, datasets, and training/extraction seeds. Results show a marginal change in accuracy of less than 1\%, while protected models resist extraction even after extended periods compared to under four hours for unprotected models. We also provide a theoretical framework to analyze the probability of defense failure, offering further insight into the effectiveness of our method.







{\selectfont{\bibliographystyle{plain}} 
\bibliography{references}}

\begin{thebibliography}{10}

\bibitem{batina2018csi}
Lejla Batina, Shivam Bhasin, Dirmanto Jap, and Stjepan Picek.
\newblock {\color{black}CSI NN: Reverse Engineering of Neural Network Architectures Through Electromagnetic Side Channel}.
\newblock In {\em 28th USENIX Security Symposium (USENIX Security 19)}, pages 515--532, 2019.

\bibitem{canales2024polynomial}
Isaac~A Canales-Mart{\'\i}nez, Jorge Ch{\'a}vez-Saab, Anna Hambitzer, Francisco Rodr{\'\i}guez-Henr{\'\i}quez, Nitin Satpute, and Adi Shamir.
\newblock {Polynomial Time Cryptanalytic Extraction of Neural Network Models}.
\newblock In {\em Annual International Conference on the Theory and Applications of Cryptographic Techniques}, pages 3--33. Springer, 2024.

\bibitem{carlini2024polynomialtimecryptanalyticextraction}
Nicholas Carlini, Jorge Chávez-Saab, Anna Hambitzer, Francisco Rodríguez-Henríquez, and Adi Shamir.
\newblock {Polynomial Time Cryptanalytic Extraction of Deep Neural Networks in the Hard-Label Setting}.
\newblock Cryptology {ePrint} Archive, Paper 2024/1580, 2024.

\bibitem{carlini2020cryptanalytic}
Nicholas Carlini, Matthew Jagielski, and Ilya Mironov.
\newblock {Cryptanalytic Extraction of Neural Network Models}.
\newblock In {\em Annual international cryptology conference}, pages 189--218. Springer, 2020.

\bibitem{chen2017targeted}
Xinyun Chen, Chang Liu, Bo~Li, Kimberly Lu, and Dawn Song.
\newblock {\color{black}Targeted Backdoor Attacks on Deep Learning Systems Using Data Poisoning}.
\newblock {\em arXiv preprint arXiv:1712.05526}, 2017.

\bibitem{cryptoeprint:2024/1403}
Yi~Chen, Xiaoyang Dong, Jian Guo, Yantian Shen, Anyu Wang, and Xiaoyun Wang.
\newblock {Hard-Label Cryptanalytic Extraction of Neural Network Models}.
\newblock Cryptology {ePrint} Archive, Paper 2024/1403, 2024.

\bibitem{fefferman1994reconstructing}
Charles Fefferman et~al.
\newblock Reconstructing a neural net from its output.
\newblock {\em Revista Matem{\'a}tica Iberoamericana}, 10(3):507--556, 1994.

\bibitem{foerster2024beyond}
Hanna Foerster, Robert Mullins, Ilia Shumailov, and Jamie Hayes.
\newblock {Beyond Slow Signs in High-fidelity Model Extraction}.
\newblock In A.~Globerson, L.~Mackey, D.~Belgrave, A.~Fan, U.~Paquet, J.~Tomczak, and C.~Zhang, editors, {\em Advances in Neural Information Processing Systems}, volume~37, pages 19496--19522. Curran Associates, Inc., 2024.

\bibitem{weightextract_sidechannel}
Cheng Gongye, Yunsi Fei, and Thomas Wahl.
\newblock {Reverse-Engineering Deep Neural Networks Using Floating-Point Timing Side-Channels}.
\newblock In {\em 2020 57th ACM/IEEE Design Automation Conference (DAC)}, pages 1--6. IEEE, 2020.

\bibitem{horvath2023barracuda}
Peter Horvath, Lukasz Chmielewski, Leo Weissbart, Lejla Batina, and Yuval Yarom.
\newblock {BarraCUDA: Edge GPUs do Leak DNN Weights}.
\newblock {\em arXiv preprint arXiv:2312.07783}, 2023.

\bibitem{jagielski2020high}
Matthew Jagielski, Nicholas Carlini, David Berthelot, Alex Kurakin, and Nicolas Papernot.
\newblock {\color{black}High Accuracy and High Fidelity Extraction of Neural Networks}.
\newblock In {\em 29th USENIX security symposium (USENIX Security 20)}, pages 1345--1362, 2020.

\bibitem{kurian2025tpuxtract}
Ashley Kurian, Anuj Dubey, Ferhat Yaman, and Aydin Aysu.
\newblock {TPUXtract: An Exhaustive Hyperparameter Extraction Framework}.
\newblock {\em IACR Transactions on Cryptographic Hardware and Embedded Systems}, 2025(1):78--103, 2025.

\bibitem{Lowd2005AdversarialL}
Daniel Lowd and Christopher Meek.
\newblock Adversarial learning.
\newblock In {\em Knowledge Discovery and Data Mining}, 2005.

\bibitem{martinelli2024expandandclusterparameterrecoveryneural}
Flavio Martinelli, Berfin Simsek, Wulfram Gerstner, and Johanni Brea.
\newblock {Expand-and-Cluster: Parameter Recovery of Neural Networks}, 2024.

\bibitem{mazeika2022steeradversarytargetedefficient}
Mantas Mazeika, Bo~Li, and David Forsyth.
\newblock {How to Steer Your Adversary: Targeted and Efficient Model Stealing Defenses with Gradient Redirection}, 2022.

\bibitem{milli2019model}
Smitha Milli, Ludwig Schmidt, Anca~D Dragan, and Moritz Hardt.
\newblock {Model Reconstruction from Model Explanations}.
\newblock In {\em Proceedings of the Conference on Fairness, Accountability, and Transparency}, pages 1--9, 2019.

\bibitem{orekondy2019knockoff}
Tribhuvanesh Orekondy, Bernt Schiele, and Mario Fritz.
\newblock {Knockoff Nets: Stealing Functionality of Black-Box Models}.
\newblock In {\em Proceedings of the IEEE/CVF conference on computer vision and pattern recognition}, pages 4954--4963, 2019.

\bibitem{orekondy2020predictionpoisoningdefensesdnn}
Tribhuvanesh Orekondy, Bernt Schiele, and Mario Fritz.
\newblock {Prediction Poisoning: Towards Defenses Against DNN Model Stealing Attacks}, 2020.

\bibitem{papernot2017practical}
Nicolas Papernot, Patrick McDaniel, Ian Goodfellow, Somesh Jha, Z~Berkay Celik, and Ananthram Swami.
\newblock {Practical Black-Box Attacks against Machine Learning}.
\newblock In {\em Proceedings of the 2017 ACM on Asia conference on computer and communications security}, pages 506--519, 2017.

\bibitem{rolnick2020reverse}
David Rolnick and Konrad Kording.
\newblock {Reverse-Engineering Deep ReLU Networks}.
\newblock In {\em International conference on machine learning}, pages 8178--8187. PMLR, 2020.

\bibitem{shamir2023polynomial}
Adi Shamir, Isaac Canales-Martinez, Anna Hambitzer, Jorge Chavez-Saab, Francisco Rodrigez-Henriquez, and Nitin Satpute.
\newblock {Polynomial Time Cryptanalytic Extraction of Neural Network Models}.
\newblock {\em arXiv preprint arXiv:2310.08708}, 2023.

\bibitem{7958568}
Reza Shokri, Marco Stronati, Congzheng Song, and Vitaly Shmatikov.
\newblock {Membership Inference Attacks Against Machine Learning Models}.
\newblock In {\em 2017 IEEE symposium on security and privacy (SP)}, pages 3--18. IEEE, 2017.

\bibitem{DBLP:journals/tches/ShuklaABMM23}
Shubhi Shukla, Manaar Alam, Sarani Bhattacharya, Pabitra Mitra, and Debdeep Mukhopadhyay.
\newblock {"Whispering MLaaS": Exploiting Timing Channels to Compromise User Privacy in Deep Neural Networks}.
\newblock {\em IACR Transactions on Cryptographic Hardware and Embedded Systems}, pages 587--613, 2023.

\bibitem{szentannai2020preventing}
K{\'a}lm{\'a}n Szentannai, Jalal Al-Afandi, and Andr{\'a}s Horv{\'a}th.
\newblock {Preventing Neural Network Weight Stealing via Network Obfuscation}.
\newblock In {\em Intelligent Computing: Proceedings of the 2020 Computing Conference, Volume 3}, pages 1--11. Springer, 2020.

\bibitem{tramer2016stealing}
Florian Tram{\`e}r, Fan Zhang, Ari Juels, Michael~K Reiter, and Thomas Ristenpart.
\newblock {Stealing Machine Learning Models via Prediction APIs}.
\newblock In {\em 25th USENIX security symposium (USENIX Security 16)}, pages 601--618, 2016.

\bibitem{deepem}
Honggang Yu, Haocheng Ma, Kaichen Yang, Yiqiang Zhao, and Yier Jin.
\newblock {DeepEM: Deep Neural Networks Model Recovery through EM Side-Channel Information Leakage}.
\newblock In {\em 2020 IEEE International Symposium on Hardware Oriented Security and Trust (HOST)}, pages 209--218. IEEE, 2020.

\bibitem{zhou2023modelobfuscator}
Mingyi Zhou, Xiang Gao, Jing Wu, John Grundy, Xiao Chen, Chunyang Chen, and Li~Li.
\newblock {ModelObfuscator: Obfuscating Model Information to Protect Deployed ML-based Systems}.
\newblock In {\em Proceedings of the 32nd ACM SIGSOFT International Symposium on Software Testing and Analysis}, pages 1005--1017, 2023.

\end{thebibliography}




\newpage
\section*{NeurIPS Paper Checklist}


\begin{enumerate}

\item {\bf Claims}
    \item[] Question: Do the main claims made in the abstract and introduction accurately reflect the paper's contributions and scope?
    \item[] Answer: \answerYes{} 
    \item[] Justification: We claim to develop an extraction-aware training method that defends against cryptanalytic attacks with zero overhead during inference, which is explained in Section~\ref{proposedWork}. We also claim to empirically evaluate our defense across different neural network configurations; these results are shown in Table~\ref{table_results}. We claim to provide a theoretical framework for computing the attack success probability, which is presented in Section~\ref{theory_probability}.

\item {\bf Limitations}
    \item[] Question: Does the paper discuss the limitations of the work performed by the authors?
    \item[] Answer: \answerYes{} 
    \item[] Justification: Our extraction-aware training strategy causes an accuracy change of less than 1\%. We quantify this extensively for all model settings used in prior works as shown in Table~\ref{table_results}. The limitations of our proposed countermeasure are discussed in Section~\ref{limitations}.

\item {\bf Theory assumptions and proofs}
    \item[] Question: For each theoretical result, does the paper provide the full set of assumptions and a complete (and correct) proof?
    \item[] Answer: \answerYes{} 
    \item[] Justification: A theoretical proof of the attack success probability is provided in Section~\ref{theory_probability}. The proof follows assumptions described in the threat model Section~\ref{threat_model}.

    \item {\bf Experimental result reproducibility}
    \item[] Question: Does the paper fully disclose all the information needed to reproduce the main experimental results of the paper to the extent that it affects the main claims and/or conclusions of the paper (regardless of whether the code and data are provided or not)?
    \item[] Answer: \answerYes{} 
    \item[] Justification: We provide details of the machines used for experiments—a high-performance cluster with Intel Xeon processors. Information about the datasets, model configurations, training, and extraction seeds used to evaluate our defense is included in Section~\ref{results}. We also report the impact factor ($\lambda_{\text{similarity}}$) required to secure each model. The code for reproducing the results is available at the GitHub link: 
    \newline
    [https://github.com/anonymous-123-code/anonymouscode]

\item {\bf Open access to data and code}
    \item[] Question: Does the paper provide open access to the data and code, with sufficient instructions to faithfully reproduce the main experimental results, as described in supplemental material?
    \item[] Answer: \answerYes{} 
    \item[] Justification: We provide a link to our GitHub code base:
    \newline
    [https://github.com/anonymous-123-code/anonymouscode].

\item {\bf Experimental setting/details}
    \item[] Question: Does the paper specify all the training and test details (e.g., data splits, hyperparameters, how they were chosen, type of optimizer, etc.) necessary to understand the results?
    \item[] Answer: \answerYes{} 
    \item[] Justification: All model settings, including configurations, training seeds, and extraction seeds, are reported in Table~\ref{table_results}. Our codebase includes both the original (non-protected) and secure models, along with scripts to create new models using the same training setup in the previous paper~\cite{foerster2024beyond}. All hyperparameters and optimizer choices are specified to enable full reproducibility. We adopt the same experimental settings, including hyperparameters and optimizer choices, as used in prior work~\cite{foerster2024beyond}.

\item {\bf Experiment statistical significance}
    \item[] Question: Does the paper report error bars suitably and correctly defined or other appropriate information about the statistical significance of the experiments?
    \item[] Answer: \answerYes{} 
    \item[] Justification: We follow the error reporting methodology provided in the previous work~\cite{foerster2024beyond} by providing the mean, variances for parameter extraction time, and the number of queries for multiple test models.

\item {\bf Experiments compute resources}
    \item[] Question: For each experiment, does the paper provide sufficient information on the computer resources (type of compute workers, memory, time of execution) needed to reproduce the experiments?
    \item[] Answer: \answerYes{} 
    \item[] Justification: We provide details of the machine setup used for experiments in Section~\ref{results}. All experiments were conducted on a high-performance cluster equipped with Intel Xeon processors, comprising 400 compute nodes and over 14,000 cores. Each node is provisioned with an average of 128 GB of memory.
    
\item {\bf Code of ethics}
    \item[] Question: Does the research conducted in the paper conform, in every respect, with the NeurIPS Code of Ethics \url{https://neurips.cc/public/EthicsGuidelines}?
    \item[] Answer: \answerYes{} 
    \item[] Justification: This research complies with the NeurIPS Code of Ethics. It does not involve human subjects, personal data, or sensitive content, and uses only publicly available datasets. Our proposed defense enhances model security without enabling misuse.

\item {\bf Broader impacts}
    \item[] Question: Does the paper discuss both potential positive societal impacts and negative societal impacts of the work performed?
    \item[] Answer: \answerYes{} 
    \item[] Justification: The paper defends against parameter extraction attacks, which helps prevent model stealing and other threats like input poisoning and membership inference. This is important for protecting intellectual property and avoiding sensitive data leaks in critical areas like healthcare and defense. This is discussed in Section~\ref{intro}.
    
\item {\bf Safeguards}
    \item[] Question: Does the paper describe safeguards that have been put in place for responsible release of data or models that have a high risk for misuse (e.g., pretrained language models, image generators, or scraped datasets)?
    \item[] Answer: \answerNA{} 
    \item[] Justification: Our work uses only publicly available datasets (MNIST) and does not involve any models with proprietary intellectual property. 

\item {\bf Licenses for existing assets}
    \item[] Question: Are the creators or original owners of assets (e.g., code, data, models), used in the paper, properly credited and are the license and terms of use explicitly mentioned and properly respected?
    \item[] Answer: \answerYes{} 
    \item[] Justification: We cite the original authors of the attacks we defend against and clearly specify the datasets used (MNIST), which are publicly available. 

\item {\bf New assets}
    \item[] Question: Are new assets introduced in the paper well documented and is the documentation provided alongside the assets?
    \item[] Answer: \answerYes{} 
    \item[] Justification: We provide a link to our GitHub code base:
\newline
    [https://github.com/anonymous-123-code/anonymouscode]

\item {\bf Crowdsourcing and research with human subjects}
    \item[] Question: For crowdsourcing experiments and research with human subjects, does the paper include the full text of instructions given to participants and screenshots, if applicable, as well as details about compensation (if any)? 
    \item[] Answer: \answerNA{} 
    \item[] Justification: This paper does not involve crowdsourcing nor research with human subjects.

\item {\bf Institutional review board (IRB) approvals or equivalent for research with human subjects}
    \item[] Question: Does the paper describe potential risks incurred by study participants, whether such risks were disclosed to the subjects, and whether Institutional Review Board (IRB) approvals (or an equivalent approval/review based on the requirements of your country or institution) were obtained?
    \item[] Answer: \answerNA{} 
    \item[] Justification: This paper does not involve crowdsourcing nor research with human subjects.

\item {\bf Declaration of LLM usage}
    \item[] Question: Does the paper describe the usage of LLMs if it is an important, original, or non-standard component of the core methods in this research? Note that if the LLM is used only for writing, editing, or formatting purposes and does not impact the core methodology, scientific rigorousness, or originality of the research, declaration is not required.
    \item[] Answer: \answerNA{} 
    \item[] Justification: The core method development in this research does not involve LLMs as any important, original, or non-standard components.

\end{enumerate}

\end{document}